\documentclass[letterpaper,10pt]{article}

\usepackage{amsmath}
\usepackage{amssymb}
\usepackage{psfrag}
\usepackage[dvips]{graphicx}
\usepackage{epsfig}

\newcommand{\dirac}{{\slash \negthinspace \negthinspace \negthinspace \nabla}}
\newcommand{\dd}{\textrm{d}}
\newcommand{\im}{{\mathbb{I}}{\mathrm{m}}}
\newcommand{\re}{{\mathbb{R}}{\mathrm{e}}}

\author{A.\ L\'opez-Ortega\thanks{alopezo@esfm.ipn.mx} \\ 
Departamento de F\'{\i}sica. Escuela Superior de F\'{\i}sica y Matem\'aticas. \\
Instituto Polit\'ecnico Nacional. \\
Unidad Profesional Adolfo L\'opez Mateos. Edificio 9. \\
M\'exico, D.\ F., M\'exico. \\
C.\ P.\ 07738 }

\title{Quasinormal frequencies of the Dirac field in a $D$-dimensional Lifshitz black hole}

\begin{document}

\maketitle

\begin{abstract}

In a $D$-dimensional Lifshitz black hole we calculate exactly the quasinormal frequencies of a test Dirac field in the massless and zero angular eigenvalue limits. These results are an extension of the previous calculations in which the quasinormal frequencies of the Dirac field are determined, but in four dimensions. We discuss the four-dimensional limit of our expressions for the quasinormal frequencies and compare with the previous results. We also determine whether the Dirac field has unstable modes in the $D$-dimensional Lifshitz spacetime.

KEYWORDS: Quasinormal modes; Dirac field; Lifshitz spacetimes; Classical stability

PACS: 04.40.-b; 04.30.Nk; 04.70.Bw; 04.50.Gh

\end{abstract}

\section{Introduction}
\label{s: Introduction}

We know condensed matter systems for which the space and time at the critical points (Lifshitz fixed points) show the anisotropic scale invariance
\begin{equation} \label{e: anisotropic scale}
 x_i \to \lambda^2 x_i,  \qquad t \to \lambda^{2 \hat{z}} t,
\end{equation} 
where $\hat{z} > 1$ is the critical exponent. To extend the AdS-CFT correspondence to systems with Lifshitz points, in recent times a detailed analysis is carried out of the spacetimes that at large $r$ asymptote to the so called Lifshitz metric \cite{Kachru:2008yh}, \cite{Balasubramanian:2009rx}
\begin{equation} \label{e: Lifshitz metric}
 \dd s^2 = \frac{r^{2 \hat{z}}}{l^{2 \hat{z}}} \dd t^2 - \frac{l^2}{r^2} \dd r^2 - r^2 \dd \Omega^2 ,
\end{equation}
where $\dd \Omega^2 $ denotes the line element of the $(D-2)$-dimensional plane $\mathbb{R}^{D-2}$ and $l$ is a positive constant. In this metric the parameter $\hat{z}$ coincides with the critical exponent of the formulas (\ref{e: anisotropic scale}).

Owing to the  quasinormal modes (QNM) are useful to determine relevant physical properties of the black holes \cite{Kokkotas:1999bd}--\cite{Konoplya:2011qq}, recently the quasinormal frequencies (QNF) of several asymptotic Lifshitz black holes have been computed \cite{Giacomini:2012hg}--\cite{Lopez-Ortega:2014oha}. In this work we study the damped oscillations of the Dirac field propagating in the $D$-dimensional ($D \geq 4$) asymptotic Lifshitz static black hole with $\hat{z}=2$ \cite{Balasubramanian:2009rx}, \cite{Giacomini:2012hg}
 \begin{equation} \label{e: black hole Lifshitz}
\dd s^2 = \frac{r^4}{l^4} \left( 1- \frac{r_+^2}{r^2} \right) \dd t^2 - \frac{l^2 \dd r^2}{r^2 -r_+^2} - r^2  \dd \Omega^2 ,
\end{equation} 
where $r_+$ locates the event horizon. The $D$-dimensional Lifshitz black hole (\ref{e: black hole Lifshitz}) is a solution of a Lagrangian with scalar and vector fields \cite{Balasubramanian:2009rx} or with higher curvature terms \cite{Giacomini:2012hg}.

To explore the classical stability of this $D$-dimensional Lifshitz black hole in Ref.\ \cite{Giacomini:2012hg} Giacomini, et al.\ calculate exactly the QNF of the massive Klein-Gordon field. They find that its spectrum of QNF is discrete and its QNM are stable. Furthermore the QNF depend on the spacetime dimension $D$ (see the expressions (33)  and (34) of Ref.\ \cite{Giacomini:2012hg}). Their results are extended in Refs.\ \cite{Catalan:2013eza}, \cite{Lopez-Ortega:2014oha}. Catal\'an, et al.\ \cite{Catalan:2013eza} calculate the QNF of the Dirac field propagating in the four-dimensional Lifshitz black hole (\ref{e: black hole Lifshitz}), that is, when $\dd \Omega^2$ is the line element of the two-dimensional plane $\mathbb{R}^2$. In the $D$-dimensional Lifshitz black hole (\ref{e: black hole Lifshitz}) the QNF of the electromagnetic field are calculated exactly in Ref.\ \cite{Lopez-Ortega:2014oha}. In this reference it is found that for $D=5,6,7,$ and for the scalar type electromagnetic field we need to impose a slightly different boundary condition as $r \to \infty$ to get a discrete and stable spectrum of QNF, since the usually imposed Dirichlet boundary condition lead us to a continuous spectrum of QNF with possible unstable QNM. For other values of the spacetime dimension the spectrum of QNF for the scalar type electromagnetic field is discrete and stable when we impose the Dirichlet boundary condition as $r \to \infty$. For the vector type electromagnetic field the usual Dirichlet boundary condition lead us to a discrete and stable spectrum of QNF. As for the Klein-Gordon field \cite{Giacomini:2012hg} the QNF of the electromagnetic field depend on the spacetime dimension. Hence in asymptotic Lifshitz black holes for the electromagnetic and Klein-Gordon fields their spectra of QNF depend on the spacetime dimension.

In this work our main objective is to show that the analytical results of Ref.\ \cite{Catalan:2013eza} in four dimensions (those on the exact calculation of the QNF for the massless Dirac field and for the massive Dirac field with angular eigenvalue equal to zero) can be extended to the $D$-dimensional Lifshitz black hole (\ref{e: black hole Lifshitz}) ($D \geq 4$), that is, when $\dd \Omega^2$ is the line element of the $(D-2)$-dimensional plane $\mathbb{R}^{D-2}$. These calculations allow us to study additional properties about the behavior of the Lifshitz black holes under small perturbations and using these results we examine whether the spectrum of QNF for the Dirac field depend on the spacetime dimension as for the Klein-Gordon and electromagnetic fields. We also study the classical stability of the Dirac field in the $D$-dimensional Lifshitz spacetime (\ref{e: Lifshitz metric}). 

In a curved spacetime it is well known that in some physical phenomena the fermion fields behave in a different way than the boson fields. For example, the fermion fields do not suffer superradiance when they are scattered by rotating black holes  \cite{Unruh:1974bw}--\cite{Iyer:1978du}. Thus we believe that is relevant to study the QNM of the Dirac field in asymptotically Lifshitz spacetimes.  See Refs.\ \cite{LopezOrtega:2009zx}--\cite{LopezOrtega:2010uu} to find other examples where the spectrum of QNF for the Dirac field is calculated exactly in other higher dimensional spacetimes. 

Thus in what follows for the massless Dirac field and for the massive Dirac field with angular eigenvalue equal to zero we calculate their QNF when they propagate in the $D$-dimensional Lifshitz black hole (\ref{e: black hole Lifshitz}) and determine whether the $D$-dimensional Lifshitz spacetime (\ref{e: Lifshitz metric}) is stable against Dirac perturbations. We note that the method exploited in this work to solve the Dirac equation in the $D$-dimensional Lifshitz black hole (\ref{e: black hole Lifshitz}) is different from the procedure used in Ref.\ \cite{Catalan:2013eza} for the four-dimensional case, since the procedure of the previous reference is adapted to four-dimensional spacetimes.

We organize this paper as follows.  Based on Refs.\ \cite{Gibbons:1993hg}--\cite{LopezOrtega:2009qc} in Sect.\ \ref{s: Dirac equation} we give the main results of the method that simplifies the Dirac equation to a pair of coupled partial equations when the $D$-dimensional background is maximally symmetric. In Sect.\ \ref{s: QNM Dirac Lifshitz black hole} we calculate exactly the QNF of the Dirac field in the $D$-dimensional Lifshitz black hole (\ref{e: black hole Lifshitz}). We compute exactly the QNF for the massive Dirac field with angular eigenvalue equal to zero and for the massless Dirac field with $\kappa \neq 0$ since for these two limiting cases we can solve exactly the radial equations. In Sect.\ \ref{s: modes Lifshitz spacetime} we determine whether the Dirac field has unstable modes in the $D$-dimensional Lifshitz spacetime (\ref{e: Lifshitz metric}) with $\hat{z}=2$. Finally we discuss some relevant facts in Sect.\ \ref{s: Discussion}.

\section{Dirac equation in $D$-dimensional maximally symmetric spacetimes}
\label{s: Dirac equation}

If $F$, $G$, and $H$ are functions of $r$ and here $\dd \Omega^2$ is the line element of a $(D-2)$-dimensional maximally symmetric space, then for a $D$-dimensional spacetime ($D \geq 4$) of the form
\begin{equation} \label{e: general metric}
\dd s^2 = F(r)^2 \dd t^2 -G(r)^2 \dd r^2 - H(r)^2 \dd \Omega^2,
\end{equation} 
it is known that the Dirac equation
\begin{equation} \label{e: Dirac equation}
 i \dirac \psi = m \psi ,
\end{equation} 
simplifies to the coupled system of partial differential equations \cite{Gibbons:1993hg}--\cite{LopezOrtega:2009qc} (see for example the formulas (30) in Ref.\  \cite{LopezOrtega:2009qc})
\begin{align} \label{e: Dirac equation coupled}
\partial_t \psi_1 + \frac{F}{G} \partial_r \psi_1 & =  - \left( i \kappa \frac{F}{H} +  i m F \right) \psi_2,  \\ 
 \partial_t  \psi_2 - \frac{F}{G} \partial_r  \psi_2 & =  \left( i \kappa \frac{F}{H} -  i m F \right) \psi_1,  \nonumber
\end{align} 
where $\psi_1$ and $\psi_2$ are the components of a two-dimensional spinor that depends on the coordinates $(t,r)$, and $\kappa$ are the eigenvalues of the Dirac operator on the $(D-2)$-dimensional maximally symmetric space with line element $\dd \Omega^2$, usually called the base manifold. 

For the $D$-dimensional Lifshitz black hole (\ref{e: black hole Lifshitz}) and for the $D$-dimensional Lifshitz spacetime (\ref{e: Lifshitz metric}) the symbol $\dd \Omega^2$ denotes the line element of the $(D-2)$-dimensional plane $\mathbb{R}^{D-2}$, which is a maximally symmetric space, therefore we can use the coupled system of partial differential equations (\ref{e: Dirac equation coupled}) to study the behavior of the Dirac field in these two backgrounds. Furthermore the eigenvalues $\kappa$ are equal to $\kappa = i \eta$ with $\eta \in \mathbb{R}$ \cite{Ginoux-book}. Notice that $\kappa = 0$ is an allowed eigenvalue of the Dirac operator on the base manifolds of the Lifshitz spacetime (\ref{e: Lifshitz metric}) and of the Lifshitz black hole (\ref{e: black hole Lifshitz}).

\section{Quasinormal modes of the Dirac field}
\label{s: QNM Dirac Lifshitz black hole}

Here we calculate exactly the QNF of the Dirac field propagating in the $D$-dimensional Lifshitz black hole (\ref{e: black hole Lifshitz}). First we notice that in the $D$-dimensional Lifshitz black hole (\ref{e: black hole Lifshitz}) the partial differential equations (\ref{e: Dirac equation coupled}) simplify to
\begin{align} \label{e: Dirac radial general}
 z(z^2 -1) \frac{\dd R_1}{\dd z} - i \tilde{\omega} R_1 &= -z (z^2 -1)^{1/2} \left(\frac{i \hat{\kappa}}{z} + i \tilde{m} \right) R_2,  \\ 
z(z^2 -1) \frac{\dd R_2}{\dd z} + i \tilde{\omega} R_2 &= -z (z^2 -1)^{1/2} \left(\frac{i \hat{\kappa}}{z} - i \tilde{m} \right) R_1,  \nonumber 
\end{align}
when the components $\psi_1$ and $\psi_2$ have the harmonic time dependence 
\begin{equation} \label{e: components spinor two}
 \psi_{j} = \textrm{e}^{- i \omega t} R_{j} (r),
\end{equation} 
with $j=1,2$. In Eqs.\ (\ref{e: Dirac radial general}) we define the quantities $z= r/r_+$, $\hat{\kappa}= (\kappa l)/r_+$, $\tilde{m} = m l$, and $\tilde{\omega} = (\omega l^3)/r_+^2$. In what follows, taking as a basis Eqs.\ (\ref{e: Dirac radial general}) we calculate exactly the QNF of the massless Dirac field with $\kappa \neq 0$ in Subsection \ref{ss: QNM masless}, and then we determine the QNF of the massive Dirac field with $\kappa = 0$ in Subsection \ref{ss: massive QNM}. We study these two limiting cases since we have not been able to simplify Eqs.\ (\ref{e: Dirac radial general}) when $m \neq 0$ and $\kappa \neq 0$ simultaneously.

\subsection{Massless Dirac field}
\label{ss: QNM masless}

For a classical field propagating in the Lifshitz black hole (\ref{e: black hole Lifshitz}) we define its QNM as the oscillations that satisfy the boundary conditions \cite{Giacomini:2012hg}--\cite{Lopez-Ortega:2014oha}
\begin{enumerate}
 \item[i)] They are purely ingoing near the horizon.
\item[ii)] They go to zero as $r \to \infty$.
\end{enumerate}

For the $D$-dimensional Lifshitz black hole (\ref{e: black hole Lifshitz}) in the massless limit the system of differential equations (\ref{e: Dirac radial general}) simplifies to
\begin{align} \label{e: Dirac radial massless}
 z(z^2 -1) \frac{\dd R_1}{\dd z} - i \tilde{\omega} R_1 &= - (z^2 -1)^{1/2} i\hat{\kappa} R_2,  \\ 
z(z^2 -1) \frac{\dd R_2}{\dd z} + i \tilde{\omega} R_2 &= - (z^2 -1)^{1/2} i \hat{\kappa} R_1 .  \nonumber 
\end{align}
From these we obtain the decoupled equations for the radial functions $R_1$ and $R_2$
\begin{align}
 \frac{\dd^2 R_j}{\dd z^2} + \left( \frac{1}{z} + \frac{z}{z^2-1} \right) \frac{\dd R_j}{\dd z} + \left( \frac{\tilde{\omega}^2 + i \epsilon \tilde{\omega}}{(z^2-1)^2} + \frac{\hat{\kappa}^2-\tilde{\omega}^2}{z^2 (z^2-1)} \right) R_j = 0 ,
\end{align}
with $\epsilon = 1$ ($\epsilon=-1$) for $R_1$ ($R_2$).
 Here we study in detail the radial function $R_1$ and we notice that similar results are valid for the radial function $R_2$.

Making the change of variable
\begin{equation} \label{e: change of variable u}
 u = \frac{z^2-1}{z^2},
\end{equation} 
and taking the function $R_1$ as
\begin{equation}
 R_1 = u^A (1-u)^{B+1/4} f_1, 
\end{equation} 
where the constants $A$ and $B$ are solutions of the algebraic equations
\begin{equation} 
 A^2 -\frac{A}{2} + \frac{\tilde{\omega}^2 + i\tilde{\omega}}{4} = 0, \qquad B^2 - \frac{1}{16} = 0,
\end{equation} 
we find that the function $f_1$ is a  solution of the hypergeometric differential equation \cite{Abramowitz-book}--\cite{NIST-book}
\begin{equation}\label{e: hypergeometric equation}
 u(1-u) \frac{\dd^2 f_1}{\dd u^2} + (c - (a+b+1)u) \frac{\dd f_1}{\dd u} - a b f_1 = 0,
\end{equation} 
with the parameters $a$, $b$, and $c$ equal to
\begin{align} \label{e: a b c masless}
 a &=A+B+\frac{1}{4} + \frac{\sqrt{\hat{\kappa}^2-\tilde{\omega}^2}}{2},\qquad  \\
 b &=A+B+\frac{1}{4} - \frac{\sqrt{\hat{\kappa}^2-\tilde{\omega}^2}}{2} ,\nonumber  \\
 c &= 2 A + \frac{1}{2}. \nonumber 
\end{align}
In what follows we take
\begin{equation}
 A = \frac{1}{2} - i\frac{\tilde{\omega}}{2}, \qquad \qquad   B= \frac{1}{4}.
\end{equation} 
We expect to find the same physical results for the other values of the parameters $A$ and $B$. 

If we assume that the quantity $c$ is not an integer to discard the solutions involving logarithmic terms \cite{Guo-book}, \cite{NIST-book}, then we get that the radial function $R_1$ is given by
\begin{align}
R_1 &= u^{1/2 - i\tilde{\omega}/2 } (1-u)^{1/2} \left( C_1 \, {}_{2}F_{1}(a,b;c;u) \right.  \nonumber \\
& +\left.  C_2 \,u^{1-c} {}_{2}F_{1}(a-c+1,b-c+1;2-c;u) \right) , 
\end{align} 
where $ {}_{2}F_{1}(a,b;c;u)$ denotes the hypergeometric function and $C_1$, $C_2$ are constants \cite{Abramowitz-book}--\cite{NIST-book}. We point out that in the coordinate $u$ the horizon of the Lifshitz black hole (\ref{e: black hole Lifshitz}) is located at $u=0$ and this coordinate satisfies $u \to 1$ as $r \to \infty$. Thus near the horizon ($u=0$) the previous radial function behaves as
\begin{align} \label{e: Dirac massless near horizon}
 R_1 &\approx C_1 u^{1/2 - i \tilde{\omega}/2} + C_2 u^{ i \tilde{\omega}/2} \\
 &\approx C_1 \exp(r_+^2 r_* /l^3) \exp(-i \omega r_*) + C_2 \exp( i \omega r_*) , \nonumber 
\end{align} 
where $r_*$ denotes the tortoise coordinate of the Lifshitz black hole (\ref{e: black hole Lifshitz}) and it is equal to
\begin{equation}
 r_* = \frac{l^3}{2 r_+^2} \ln (u) = \frac{l^3}{2 r_+^2} \ln \left(\frac{z^2-1}{z^2} \right) , 
\end{equation} 
that is, $r_* \in (-\infty, 0)$ for $r \geq r_+$.

Considering that we have a time dependence of the form $\exp(-i \omega t)$ (see the formula (\ref{e: components spinor two})), near the horizon the first term of the formula (\ref{e: Dirac massless near horizon}) is an ingoing wave and the second term represents an outgoing wave. Hence to have a purely ingoing wave near the horizon we must take $C_2=0$. Thus the radial function that satisfies the boundary condition of the QNM near the horizon is
\begin{align}\label{e: radial massive Kummer}
 R_1 &= C_1 u^{1/2 - i\tilde{\omega}/2 } (1-u)^{1/2} {}_{2}F_{1}(a,b;c;u) \\
 &= C_1 u^{1/2 - i\tilde{\omega}/2 } \left[ (1-u)^{1/2} \frac{\Gamma(c) \Gamma(c-a-b)}{\Gamma(c-a) \Gamma(c-b)} \, {}_2 F_1 (a,b;a+b-c+1;1-u) \right. \nonumber \\
 &\left. + \frac{\Gamma(c) \Gamma(a+b-c)}{\Gamma(a) \Gamma(b)} {}_2F_1(c-a,c-b;c+1-a-b; 1 -u) \nonumber\right] ,
\end{align}
where in the second line of the previous expression we use the Kummer formula for the hypergeometric function $ {}_{2}F_{1}(a,b;c;u)$ that for $c-a-b$ different from an integer establishes \cite{Abramowitz-book}--\cite{NIST-book} (see for example the formula (4) of Sect.\ 4.8 in Ref.\ \cite{Guo-book})
\begin{align} \label{e: Kummer property y 1-y}
&{}_2F_1(a,b;c;u) = \frac{\Gamma(c) \Gamma(c-a-b)}{\Gamma(c-a) \Gamma(c - b)} {}_2 F_1 (a,b;a +b +1-c;1-u)  \\
&+ \frac{\Gamma(c) \Gamma( a + b - c)}{\Gamma(a) \Gamma(b)} (1-u)^{c-a-b} {}_2F_1(c-a, c-b; c + 1 -a-b; 1 -u). \nonumber
\end{align}
In our case we can use the Kummer formula since $c-a-b = -1/2$, which differs from an integer.

Therefore from the expression (\ref{e: radial massive Kummer}) for the function $R_1$ we obtain that as $u \to 1$ 
\begin{equation}
 \lim_{u \to 1} R_1 = C_1 \frac{\Gamma(c) \Gamma(a+b-c)}{\Gamma(a) \Gamma(b)},
\end{equation} 
since the first term goes to zero in this limit and to satisfy the boundary condition ii) of the QNM  as $r \to \infty$ we must impose
\begin{equation}
 a = - n , \qquad \textrm{or} \qquad b = -n, \qquad n=0,1,2,\dots,
\end{equation} 
from which we find that in the $D$-dimensional Lifshitz black hole (\ref{e: black hole Lifshitz}) the QNF of the massless Dirac field are equal to
\begin{equation} \label{e: QNF Dirac massive 1}
 \omega = - \frac{ir_+^2}{l^3} \left( n + 1 + \frac{\eta^2 l^2 }{r_+^2(4 n +4)} \right),
\end{equation} 
with $\eta \in \mathbb{R}$ \cite{Ginoux-book}. Using a similar method we get that for the component $\psi_2$ its QNF are equal to
\begin{equation} \label{e: QNF Dirac massive 2}
 \omega = - \frac{ir_+^2}{l^3} \left( n + \frac{1}{2} + \frac{\eta^2 l^2 }{r_+^2(4 n +2)} \right).
\end{equation} 

For $D=4$ and making the identification $r_+^2=l^2/2$, as in Ref.\ \cite{Catalan:2013eza}, we find that the QNF (\ref{e: QNF Dirac massive 1}) and (\ref{e: QNF Dirac massive 2}) of the massless Dirac field become
\begin{equation} \label{e: QNF Catalan massless}
 \omega = -\frac{i}{l} \left(\frac{n}{2} + \frac{1}{2} + \frac{\eta^2}{2(2n +2)} \right), \qquad \omega = -\frac{i}{l} \left(\frac{n}{2} + \frac{1}{4} + \frac{\eta^2}{2(2n +1)} \right) ,
\end{equation} 
that are equal to the QNF of the Dirac field given in Ref.\ \cite{Catalan:2013eza}. Moreover we note that for the QNF (\ref{e: QNF Dirac massive 1}) and (\ref{e: QNF Dirac massive 2}) the quantity $c$ of the formulas (\ref{e: a b c masless}) is not an integer, as we assumed.

\subsection{Massive Dirac field with $\kappa = 0$}
\label{ss: massive QNM}

Another limit for which we can solve exactly the radial equations is for the massive Dirac field with angular eigenvalue equal to zero. Therefore in what follows we calculate exactly the QNF of the massive Dirac field with $\kappa = 0$ propagating in the $D$-dimensional Lifshitz black hole (\ref{e: black hole Lifshitz}). In contrast to Catal\'an, et al.\ \cite{Catalan:2013eza} that consider negative values of the mass $m$,  we study positive values of $m$. For $\kappa = 0$ the system of differential equations (\ref{e: Dirac radial general}) simplifies to
\begin{align} \label{e: Dirac radial kappa}
 z(z^2 -1) \frac{\dd R_1}{\dd z} - i \tilde{\omega} R_1 &= - z (z^2 -1)^{1/2} i \tilde{m} R_2,  \\ 
z(z^2 -1) \frac{\dd R_2}{\dd z} + i \tilde{\omega} R_2 &=  z (z^2 -1)^{1/2} i \tilde{m} R_1 ,  \nonumber 
\end{align}
from which we obtain the following decoupled equations for the radial functions $R_1$ and $R_2$
\begin{align}
 \frac{\dd^2 R_j}{\dd z^2} +  \frac{z}{z^2-1} \frac{\dd R_j}{\dd z} + \left(\frac{\tilde{\omega}^2 -i\epsilon \tilde{\omega} }{z^2} + \frac{\tilde{\omega}^2 + i\epsilon \tilde{\omega}}{(z^2-1)^2} - \frac{\tilde{\omega}^2 -i \epsilon \tilde{\omega} + \tilde{m}^2}{z^2-1} \right) R_j = 0 ,
\end{align}
where $\epsilon$ takes the same values for $R_1$ and $R_2$ that in the previous subsection.

Here we study in detail the radial function $R_1$ (similar results are valid for the function $R_2$). Making the change of variable (\ref{e: change of variable u}) and taking $R_1$ as
\begin{equation}
 R_1 = u^A (1-u)^{B} f_1, 
\end{equation} 
with the quantities $A$ and $B$ being solutions of the algebraic equations 
\begin{equation}
 A^2 -\frac{A}{2} + \frac{\tilde{\omega}^2 + i\tilde{\omega}}{4} = 0, \qquad B^2 - \frac{\tilde{m}^2}{4} = 0,
\end{equation} 
we find that the function $f_1$ is a solution of the hypergeometric differential equation (\ref{e: hypergeometric equation}) with parameters $a$, $b$, and $c$ equal to \cite{Abramowitz-book}--\cite{NIST-book}
\begin{align} \label{e: a b c Dirac kappa}
 a &=A+B+\frac{1}{4} + \frac{\sqrt{-\tilde{\omega}^2 + i \tilde{\omega} + 1/4 }}{2},\qquad  \\
 b &=A+B+\frac{1}{4} -  \frac{\sqrt{-\tilde{\omega}^2 + i \tilde{\omega} + 1/4 }}{2} , \nonumber \\
 c &= 2 A +\frac{1}{2} . \nonumber 
\end{align}
In what follows we take
\begin{equation}
 A = \frac{1}{2} - i\frac{\tilde{\omega}}{2}, \qquad \qquad   B= \frac{\tilde{m}}{2}.
\end{equation} 
We expect to get similar results for the other values of the constants $A$ and $B$.

If the parameter $c$ is not an integer, then the radial function $R_1$ is given by \cite{Guo-book}, \cite{NIST-book}
\begin{align}
R_1 &= u^{1/2 - i\tilde{\omega}/2 } (1-u)^{\tilde{m}/2} \left( C_1 \, {}_{2}F_{1}(a,b;c;u) \right. \nonumber \\ 
&+\left. C_2 \, u^{1-c} {}_{2}F_{1}(a-c+1,b-c+1;2-c;u) \right) , 
\end{align}
with $C_1$ and $C_2$ constants. Near the horizon of the Lifshitz black hole (\ref{e: black hole Lifshitz}) ($u=0$) we find that the previous radial function behaves as
\begin{align} \label{e: Dirac kappa near horizon}
 R_1  \approx C_1 \exp(r_+^2 r_* /l^3) \exp(-i \omega r_*) + C_2 \exp( i \omega r_*) . 
\end{align} 
Since we assume a time dependence of the form $\exp(-i \omega t)$ we see that in the previous approximation for the radial function $R_1$ the first term is an ingoing wave, whereas the second term is an outgoing wave. Hence to get a purely ingoing wave near the horizon of the Lifshitz black hole (\ref{e: black hole Lifshitz}) we must impose $C_2= 0$.

Therefore the radial function satisfying the boundary condition i) of the QNM is equal to
\begin{align}
 R_1 &= C_1 u^{1/2 - i\tilde{\omega}/2 } (1-u)^{ \tilde{m}/2} {}_{2}F_{1}(a,b;c;u) \\
 &= C_1 u^{1/2 - i\tilde{\omega}/2 } \left[ (1-u)^{\tilde{m} /2} \frac{\Gamma(c) \Gamma(c-a-b)}{\Gamma(c-a) \Gamma(c-b)} \, {}_2 F_1 (a,b;a+b+1-c;1-u) \right. \nonumber \\
 &\left. + (1-u)^{ -\tilde{m} /2} \frac{\Gamma(c) \Gamma(a+b-c)}{\Gamma(a) \Gamma(b)} {}_2F_1(c-a,c-b;c+1-a-b; 1 -u) \nonumber\right] ,
\end{align}
where in the second line of the previous equation we use Kummer's formula for the hypergeometric function (\ref{e: Kummer property y 1-y}), since the quantity $c-a-b= -\tilde{m}$ is not an integer \cite{Abramowitz-book}--\cite{NIST-book}. From the last expression for the radial function $R_1$, in the limit $u \to 1$ we note that the first term goes to zero, whereas the second diverges in this limit. Therefore to satisfy the boundary condition ii) of the QNM we must impose 
\begin{equation}
 a = - n ,\qquad \textrm{or} \qquad b = -n , \qquad n=0,1,2,3,\dots
\end{equation} 
Using the values for the parameters $a$ and $b$ of the formulas (\ref{e: a b c Dirac kappa}) we obtain that in the $D$-dimensional Lifshitz black hole (\ref{e: black hole Lifshitz}) the QNF of the massive Dirac field with $\kappa = 0$ are equal to
\begin{equation} \label{e: QNF Dirac kappa}
 \omega = - \frac{ir_+^2}{l^3} \left( n + \frac{1}{2} + \frac{\tilde{m}}{2} \right).
\end{equation}

For the radial function $R_2$ a similar method also gives the previous QNF. If we make $r_+^2=l^2/2$, as previously, for $D=4$ the QNF (\ref{e: QNF Dirac kappa}) coincide with those calculated in Ref.\ \cite{Catalan:2013eza}, that is, the formula (\ref{e: QNF Dirac kappa}) produces the values reported in the previous reference, except that our expression (\ref{e: QNF Dirac kappa}) gives the additional QNF $-(i/l)(1/4+ml/4)$ for the component $\psi_2$ of Ref.\ \cite{Catalan:2013eza}. Furthermore, for the QNF (\ref{e: QNF Dirac kappa}) we get that the parameter $c$ of the formulas (\ref{e: a b c Dirac kappa}) is not an integer, as we assumed.

\section{Modes of the Dirac field in the Lifshitz spacetime}
\label{s: modes Lifshitz spacetime}

To extend the results of the previous section here we determine the modes of the Dirac field in the $D$-dimensional Lifshitz spacetime (\ref{e: Lifshitz metric}). In a similar way to the electromagnetic field \cite{Lopez-Ortega:2014oha}, in the Lifshitz spacetime (\ref{e: Lifshitz metric}) the modes of the Dirac field must satisfy the boundary conditions:
\begin{enumerate}
 \item[1)] The modes go to zero as $r \to \infty$.
 \item[2)] The modes are regular at $r=0$.
\end{enumerate}
Our objective is to determine for the Dirac field the existence of unstable modes that satisfy the previous boundary conditions, that is, for the Dirac field propagating in the $D$-dimensional Lifshitz spacetime (\ref{e: Lifshitz metric}) we are looking for modes whose amplitudes increase with the time and that fulfill the boundary conditions 1) and 2).

For the Dirac field propagating in the $D$-dimensional Lifshitz spacetime (\ref{e: Lifshitz metric}) the coupled system of differential equations (\ref{e: Dirac equation coupled}) simplify to
\begin{align} \label{e: Dirac radial Lifshitz general}
 y^{\hat{z}+1} \frac{\dd R_1}{\dd y} - i \hat{\omega} R_1 &= - y^{\hat{z}} \left(\frac{i \kappa}{y} + i \hat{m} \right) R_2,  \\ 
 y^{\hat{z}+1} \frac{\dd R_2}{\dd y} + i \hat{\omega} R_2 &= - y^{\hat{z}} \left(\frac{i \kappa}{y} - i \hat{m} \right) R_1,  \nonumber 
\end{align}
when we take the components $\psi_1$ and $\psi_2$ as in the formula (\ref{e: components spinor two}) and we define the quantities $y=r/l$, $\hat{\omega} = \omega l$, and $\hat{m} = m l$. Since we have not been able to simplify in an appropriate form the system of differential equations (\ref{e: Dirac radial Lifshitz general}) when $m \neq 0$ and $\kappa \neq 0$ simultaneously, in a similar way to the Lifshitz black hole (\ref{e: black hole Lifshitz}), in what follows we calculate the modes of the massless Dirac field with $\kappa \neq 0$ and of the massive Dirac field with $\kappa = 0$.

\subsection{Massless Dirac field ($\kappa \neq 0$)}

For the Dirac field propagating in the $D$-dimensional Lifshitz spacetime (\ref{e: Lifshitz metric}), in the massless limit we get that the coupled system of differential equations (\ref{e: Dirac radial Lifshitz general}) simplifies to
\begin{align} \label{e: Dirac radial Lifshitz massless}
 y \frac{\dd R_1}{\dd y} - \frac{i \hat{\omega}}{y^{\hat{z}}} R_1 &= - \frac{i \kappa}{y} R_2,  \\ 
 y \frac{\dd R_2}{\dd y} + \frac{i \hat{\omega}}{y^{\hat{z}}} R_2 &= - \frac{i \kappa}{y} R_1,  \nonumber 
\end{align}
from which we obtain for the functions $R_1$ and $R_2$ the decoupled equations 
\begin{equation} \label{e: radial Lifshitz general}
 \frac{\dd^2 R_j}{\dd y^2} + \frac{2}{y} \frac{\dd R_j}{\dd y} + \left( \frac{\hat{\omega}^2}{y^{2 \hat{z} + 2}} + \frac{i \hat{\omega}(z-1) \epsilon }{y^{\hat{z} + 2}} + \frac{\kappa^2}{y^4} \right) R_j = 0,
\end{equation} 
with $j=1,2,$ and $\epsilon = 1$ for $R_1$ whereas $\epsilon = -1$ for $R_2$, as previously. In a similar way to the $D$-dimensional Lifshitz black hole (\ref{e: black hole Lifshitz}) we restrict to the case $\hat{z} = 2$, since for this value of the critical exponent we can solve exactly the radial equations (\ref{e: radial Lifshitz general}). Thus for $\hat{z} = 2$ we find that the differential equations (\ref{e: radial Lifshitz general}) take the form 
\begin{equation} \label{e: radial Lifshitz general x variable}
  \frac{\dd^2 R_j}{\dd x^2} + (\kappa^2 + \epsilon i \hat{\omega} ) R_j + \hat{\omega}^2 x^2 R_j = 0,
\end{equation} 
where we define the coordinate $x$ by
\begin{equation} \label{e: x definition}
 x = \frac{1}{y} .
\end{equation} 

Making the ansatz 
\begin{equation} \label{e: ansatz R}
 R_j = \textrm{e}^{i \hat{\omega} x^2 / 2}  \hat{R}_j ,
\end{equation} 
we obtain that the functions $ \hat{R}_j$ must be a solution of the differential equation
\begin{equation} \label{e: radial v massless}
 v \frac{\dd^2 \hat{R}_j}{\dd v^2} + \left(\frac{1}{2} - v \right) \frac{\dd \hat{R}_j}{\dd v}  - \frac{i \hat{\omega} (\epsilon +1) - \eta^2 }{4 i \hat{\omega} } \hat{R}_j = 0,
\end{equation} 
where we use that $\kappa = i \eta$, as previously \cite{Ginoux-book}, and we define the variable $v$ by
\begin{equation} \label{e: definition v}
 v = - i \hat{\omega} x^2 .
\end{equation}  
We notice that the differential equations (\ref{e: radial v massless}) for the functions $\hat{R}_j$ are confluent hypergeometric differential equations \cite{Guo-book}, \cite{NIST-book}
\begin{equation} \label{e: confluent hypergeometric equation}
 v \frac{\dd^2 f }{\dd v^2} + (b -v)\frac{\dd f}{\dd v} - a f = 0,
\end{equation} 
with parameters
\begin{equation}
 a_j = - \frac{\eta^2 }{4 i \hat{\omega}} + \frac{\epsilon +1}{4} , \qquad \quad b_j=\frac{1}{2}.
\end{equation} 

If $C_{1j}$ and $C_{2j}$ are constants, the solutions of the differential equations (\ref{e: radial Lifshitz general x variable}) are \cite{NIST-book}
\begin{equation} \label{e: radial massless Lifshitz general solution}
 R_j = C_{1j} \textrm{e}^{-v/2} U (a_j,b_j; v) + C_{2j} \textrm{e}^{v/2} U(b_j-a_j,b_j;\textrm{e}^{-i \pi} v), 
\end{equation} 
where $U (a,b; v)$ denotes the Tricomi solution of the confluent hypergeometric differential equation (\ref{e: confluent hypergeometric equation}) \cite{NIST-book}. Since we are looking for unstable modes, in what follows we assume that the imaginary parts of the frequencies fulfill $\im (\omega) > 0$, and hence from the formula (\ref{e: definition v}) we get that $\re (v) > 0$. Notice that we choose a time dependence $\textrm{exp}(-i \omega t) $ (see the formulas (\ref{e: components spinor two})) and for $\im (\omega) > 0$ the amplitude of the Dirac field increases with the time, that is, for $\im (\omega) > 0$ the modes are unstable.

For $v \to \infty$ ($r \to 0$) we find that the functions $R_j$ of the formula (\ref{e: radial massless Lifshitz general solution}) behave as
\begin{equation} \label{e: approximation R massless}
 R_j \approx \frac{C_{1j}}{\textrm{e}^{v/2} v^{a_j} } + C_{2j} \textrm{e}^{v/2} v^{a_j-b_j},
\end{equation} 
where we use that the Tricomi solution satisfies  \cite{NIST-book}
\begin{equation} \label{e: Tricomi infinity}
 U (a,b; v) \approx v^{-a} ,
\end{equation} 
as $v \to \infty$. Owing to $\re (v) > 0$ we find that the first term in the formula (\ref{e: approximation R massless}) goes to zero as $v \to \infty$, whereas the second term diverges in this limit. Hence to fulfill the boundary condition 2) we must take $C_{2j} = 0$, and the radial functions $R_j$ simplify to
\begin{equation} \label{e: radial massless Lifshitz}
 R_j = C_{1j} \textrm{e}^{-v/2} U (a_j,b_j; v).
\end{equation} 

It is convenient to recall that when the quantity $b$ is different from an integer the Tricomi solution $U (a,b; v)$ satisfies \cite{NIST-book}
\begin{equation} \label{e: Tricomi property}
 U (a,b; v) = \frac{\Gamma(1-b)}{\Gamma(a-b+1)} {}_{1}F_{1}(a,b;v) + \frac{\Gamma(b-1)}{\Gamma(a)} v^{1-b} {}_{1}F_{1}(a-b+1,2-b;v),
\end{equation} 
where $ {}_{1}F_{1}(a,b;v)$ is the confluent hypergeometric function \cite{Guo-book}, \cite{NIST-book}. Using this property of the Tricomi solution we write the radial functions (\ref{e: radial massless Lifshitz}) as
\begin{align} \label{e: massless solution v 1}
 R_j &=  C_{1j} \left[ \frac{\Gamma(1/2) \textrm{e}^{-v/2} }{\Gamma(a_j + 1/2)} {}_{1}F_{1}(a_j,b_j;v) \right. \nonumber \\
&+ \left.  \frac{\Gamma(-1/2)}{\Gamma(a_j)}  \textrm{e}^{-v/2} v^{1/2} {}_{1}F_{1}(a_j-b_j+1,3/2;v) \right].
\end{align} 
In the limit $v \to 0$ ($r \to \infty$) we get that the previous functions behave in the form 
\begin{equation}
 \lim_{v \to 0} R_j \approx \frac{\Gamma(1/2)}{\Gamma(a_j + 1/2)} .
\end{equation} 
Hence to satisfy the boundary condition 1) we must impose
\begin{equation}
 a_j + \frac{1}{2} = -n, \qquad \qquad n=0,1,2,3,\dots,
\end{equation} 
from which we obtain that the frequencies of the modes that fulfill the boundary conditions 1) and 2) are equal to
\begin{equation} \label{e: frequecies massless pure}
 \hat{\omega}_j = - i \frac{\eta^2}{4} \frac{1}{n + (\epsilon +1)/4 + 1/2} .
\end{equation} 

Considering that for $\epsilon= 1$ and $\epsilon= -1$ it is true that
\begin{equation}
 \frac{\epsilon +1}{4} + \frac{1}{2} > 0,
\end{equation} 
we obtain that for the frequencies (\ref{e: frequecies massless pure}) their imaginary parts satisfy $\im (\hat{\omega}_j) < 0$, and we notice that this fact contradicts our assumption that the frequencies of the modes fulfill $\im (\omega) > 0$. Therefore we do not find unstable modes satisfying the boundary conditions 1) and 2) and we infer that the modes of the massless Dirac field with $\kappa \neq 0$ are stable in the $D$-dimensional Lifshitz spacetime (\ref{e: Lifshitz metric}).

\subsection{Massive Dirac field with $\kappa = 0$}

As for the $D$-dimensional Lifshitz black hole (\ref{e: black hole Lifshitz}), in the $D$-dimensional Lifshitz spacetime (\ref{e: Lifshitz metric}) with $\hat{z} = 2$ we can solve exactly the radial equations (\ref{e: Dirac radial Lifshitz general}) of the massive Dirac field  in the limit when the angular eigenvalue goes to zero. It is convenient to recall that $\kappa=0$ is an allowed eigenvalue for the Dirac operator on the base manifold of the Lifshitz spacetime (\ref{e: Lifshitz metric}) \cite{Ginoux-book}. In this limit, from Eqs.\ (\ref{e: Dirac radial Lifshitz general}) we obtain the following decoupled equations for the functions $R_1$ and $R_2$
\begin{equation}
  \frac{\dd^2 R_j}{\dd y^2} + \frac{1}{y} \frac{\dd R_j}{\dd y} + \left( \frac{\hat{\omega}^2}{y^6} + \frac{2 i \hat{\omega} \epsilon }{y^{4}} - \frac{\hat{m}^2}{y^2} \right) R_j = 0,
\end{equation} 
where the quantities $j$ and $\epsilon$ take the same values that in the previous sections. Making the change of variable (\ref{e: x definition}) and taking the functions $R_j$ as in the formula (\ref{e: ansatz R}), in this case we find that the functions $\hat{R}_j$ of this expression must be solutions of the differential equations 
\begin{equation} \label{e: radial v angular 1}
  \frac{\dd^2 \hat{R}_j}{\dd v^2} + \left(\frac{1}{v} - 1 \right) \frac{\dd \hat{R}_j}{\dd v}  - \left( \frac{ (\epsilon +1) }{2 v } + \frac{\hat{m}^2}{4v^2} \right) \hat{R}_j = 0,
\end{equation} 
where we use the variable $v$ of the formula (\ref{e: definition v}).

Proposing that the functions  $\hat{R}_j$ take the form
\begin{equation} \label{e: ansatz radial massive}
 \hat{R}_j = v^{A_j} \check{R}_j,
\end{equation} 
with the quantities $A_j$ being solutions of 
\begin{equation}
 A_j^2 - \frac{\hat{m}^2}{4} = 0 ,
\end{equation} 
and substituting the expression (\ref{e: ansatz radial massive}) into Eq.\ (\ref{e: radial v angular 1}), we find that the functions $ \check{R}_j$ must solve the differential equations\footnote{It is convenient to note that in Eqs.\  (\ref{e: radial v angular 2}) the frequency $\hat{\omega}$ does not appear explicitly and it is contained only in the variable $v$ (see the definition (\ref{e: definition v})).}
\begin{equation} \label{e: radial v angular 2}
  v \frac{\dd^2 \check{R}_j}{\dd v^2} + \left( 2A_j + 1 - v \right) \frac{\dd \check{R}_j}{\dd v}  - \left( \frac{1+\epsilon}{2} + A_j \right) \check{R}_j = 0 .
\end{equation} 
As in the previous subsection these are confluent hypergeometric differential equations (\ref{e: confluent hypergeometric equation}) with parameters
\begin{equation}
 a_j = A_j + \frac{1+ \epsilon}{2}, \qquad \quad  b_j= 2 A_j + 1.
\end{equation} 
In what follows we take $A_j = \hat{m}/2 $ and hence the previous quantities are equal to
\begin{equation}
 a_j =  \frac{\hat{m} + (1+\epsilon)}{2} , \qquad \quad b_j= 1 + \hat{m}.
\end{equation} 

Thus the radial functions $R_j$ are
\begin{equation} \label{e: radial angular solution}
 R_j = \textrm{e}^{-v/2} v^{\hat{m}/2}\left(  C_{1j}  U (a_j,b_j; v) + C_{2j} \textrm{e}^{v} U(b_j-a_j,b_j;\textrm{e}^{-i \pi} v) \right) , 
\end{equation} 
where $C_{1j}$ and $C_{2j}$ are constants, as previously. In what follows we assume that the imaginary parts of the frequencies satisfy $\im (\omega) > 0$, as in the previous subsection, and using the property (\ref{e: Tricomi infinity}) of the Tricomi solution, we find that the second term of the radial functions (\ref{e: radial angular solution}) diverges as $v \to \infty$. Therefore to get a well behaved solution as $v \to \infty$ we take $C_{2j}=0$ and the radial functions that satisfy the boundary condition 2) are equal to
\begin{equation}
 R_j = C_{1j} \textrm{e}^{-v/2} v^{\hat{m}/2} U (a_j,b_j; v).
\end{equation} 
Taking into account the property (\ref{e: Tricomi property}) of the Tricomi solution we obtain that the previous radial functions transform into
\begin{align} \label{e: angular solution v 1}
 R_j &=  C_{1j} \textrm{e}^{-v/2} v^{\hat{m}/2} \left[ \frac{\Gamma(1-b_j)}{\Gamma(a_j-b_j+1)}   {}_{1}F_{1}(a_j,b_j;v) \right. \nonumber \\
&+ \left.  \frac{\Gamma(b_j-1)}{\Gamma(a_j)} v^{1-b_j}  {}_{1}F_{1}(a_j-b_j+1,2-b_j;v) \right] ,
\end{align} 
that in the limit $v \to 0$ behave as
\begin{equation}
 R_j \approx \frac{\Gamma(1 - b_j) }{\Gamma(a_j -b_j + 1)} \frac{v^{\hat{m}/2}}{\textrm{e}^{v/2}} + \frac{\Gamma(b_j-1)}{\Gamma(a_j)} \frac{1}{\textrm{e}^{v/2} v^{\hat{m}/2}} .  
\end{equation} 

From this expression we notice that in the limit $v \to 0$ the first term goes to zero, whereas the second term diverges. Hence to fulfill the boundary condition 1) we must impose
\begin{equation}
 a_j = -n, \qquad \qquad n=0,1,2,3,\dots,
\end{equation}
form which we obtain
\begin{equation} \label{e: condition angular}
 \frac{\hat{m}}{2} + \frac{1 + \epsilon}{2} = -n.
\end{equation} 
Since in the previous formulas for $\epsilon=1$ and $\epsilon=-1$ the left hand side is positive we can not satisfy the conditions (\ref{e: condition angular}) and hence for the massive Dirac field with angular eigenvalue equal to zero we do not find modes with $\im (\omega) > 0$ that fulfill the boundary conditions 1) and 2). Therefore for this field we do not obtain unstable modes satisfying the boundary conditions 1) and 2).

\section{Discussion}
\label{s: Discussion}

From the expressions (\ref{e: QNF Dirac massive 1}) and (\ref{e: QNF Dirac massive 2}) for the QNF of the massless Dirac field and from the formula (\ref{e: QNF Dirac kappa}) for the QNF of the massive Dirac field with $\kappa = 0$ we find that their imaginary parts satisfy $\im(\omega) < 0$ and considering that we take a time dependence of the form $\exp(- i \omega t)$ we get that in the $D$-dimensional Lifshitz black hole (\ref{e: black hole Lifshitz}) the QNM of the massless Dirac field with $\kappa \neq 0$ and of the massive Dirac field with $\kappa = 0$ are stable since their amplitudes decay as the time increases. Thus when we impose the Dirichlet boundary condition ii) as $r \to \infty$, for the Dirac field we find a discrete spectrum of QNF for all $D \geq 4$ and its QNM are stable, as for the Klein-Gordon field \cite{Giacomini:2012hg}. Notice that for the electromagnetic field of scalar type propagating in the $D$-dimensional Lifshitz black hole (\ref{e: black hole Lifshitz}) we need to impose a slightly different boundary condition as $r \to \infty$ when $D=5,6,7$ \cite{Lopez-Ortega:2014oha}.  

Since for the Dirac field propagating in the $D$-dimensional Lifshitz black hole (\ref{e: black hole Lifshitz}) from the expressions (\ref{e: QNF Dirac massive 1}), (\ref{e: QNF Dirac massive 2}), and (\ref{e: QNF Dirac kappa}) we find that  its QNF do not depend explicitly on the spacetime dimension, and hence the QNM of the Dirac field behave in a different way than those of the electromagnetic and Klein-Gordon fields.

In a similar way to the electromagnetic field \cite{Lopez-Ortega:2014oha}, we also show that in the $D$-dimensional Lifshitz spacetime (\ref{e: Lifshitz metric}) with $\hat{z} = 2$ the massless Dirac field with $\kappa \neq 0$ and the massive Dirac field with $\kappa = 0$ do not have unstable modes that satisfy the boundary conditions 1) and 2) of Sect.\ \ref{s: modes Lifshitz spacetime}. Thus we expect that the $D$-dimensional Lifshitz spacetime (\ref{e: Lifshitz metric}) with $\hat{z} = 2$ is linearly stable against Dirac perturbations.

Doubtless, for the $D$-dimensional Lifshitz black hole (\ref{e: black hole Lifshitz}) and the $D$-di\-men\-sion\-al Lifshitz spacetime (\ref{e: Lifshitz metric}) the generalization of our results for the Dirac field with $m \neq 0$ and $\kappa \neq 0$ simultaneously deserves further research.

We recall that in the two-dimensional Witten black hole the QNF of the massive Dirac field are equal to \cite{LopezOrtega:2011sc} (see the formulas (61) and (62) in Ref.\ \cite{LopezOrtega:2011sc})
\begin{align}
 \omega_1 = -i\left(\frac{n}{2} + \frac{1}{2} - \frac{m_w^2}{(2n +2)} \right) ,\qquad &\omega_1 = -i \left(\frac{n}{2} + \frac{1}{4} - \frac{m_w^2}{(2n +1)} \right), \nonumber \\ 
\omega_2 =  -i \left(\frac{n}{2} + \frac{1}{4} - \frac{m_w^2}{(2n +1)}   \right) ,\qquad &\omega_2 = -i \left(\frac{n}{2}  - \frac{m_w^2}{2n}  \right), 
\end{align}
where $m_w$ is the mass of the Dirac field in the two-dimensional black hole and $n=1,2,3,\dots$, in the second set of QNF $\omega_2$. If in the values of the previous QNF $\omega_1$ we make the identification $m_w = i \eta / \sqrt{2}$ and we also divide by $l$ the whole expressions of the QNF $\omega_1$, then we obtain the QNF (\ref{e: QNF Catalan massless}) of the massless Dirac field in the Lifshitz black hole (\ref{e: black hole Lifshitz}). Something similar happens with the QNF $\omega_2$ of the two-dimensional Witten black hole.  We do not know an explication for this fact, but it points out to some connection between these two backgrounds.

In Ref.\ \cite{Emparan:2014cia} Emparan and Tanabe show that for a large class of black holes, in the limit $D \to \infty$ their QNF behave as 
\begin{equation} \label{e: Emparan QNF result}
 \omega_D = \left[ \frac{D}{2} + K - \left(\frac{e^{i \pi}}{2} \left( \frac{D}{2}+ K \right) \right)^{1/3} \alpha_p \right]\frac{1}{r_+},
\end{equation} 
where $K$ is the angular momentum number, $r_+$ is the radius of the event horizon, and the quantities $-\alpha_p$ are the zeroes of the Airy function. Emparan and Tanabe proposes that the behavior for the QNF given in the formula (\ref{e: Emparan QNF result}) would be true for other black holes, but as shown in Ref.\ \cite{Lopez-Ortega:2014oha} for the electromagnetic and Klein-Gordon fields propagating in the $D$-dimensional Lifshitz black hole (\ref{e: black hole Lifshitz}) the expression (\ref{e: Emparan QNF result}) does not describe the behavior of their QNF in the limit $D \to \infty$. From our expressions (\ref{e: QNF Dirac massive 1}), (\ref{e: QNF Dirac massive 2}), and (\ref{e: QNF Dirac kappa}) for the QNF of the Dirac field and taking into account that they do not depend on the spacetime dimension $D$, we infer that for the Dirac field propagating in the Lifshitz black hole (\ref{e: black hole Lifshitz}) the formula (\ref{e: Emparan QNF result}) does not produce the behavior of its QNF in the limit $D \to \infty$. For example, the QNF (\ref{e: Emparan QNF result}) are complex numbers with real part different from zero, but in the $D$-dimensional Lifshitz black hole the QNF (\ref{e: QNF Dirac massive 1}), (\ref{e: QNF Dirac massive 2}), and (\ref{e: QNF Dirac kappa}) of the Dirac field are purely imaginary. Furthermore the QNF of the Dirac field in the $D$-dimensional Lifshitz black hole (\ref{e: black hole Lifshitz}) depend on the radius of the horizon as $r_+^2$, whereas in the formula (\ref{e: Emparan QNF result}) the dependence on the radius of the horizon is in the form $1/r_+$. These facts suggest that in the limit $D \to \infty$ the behavior of the QNF is not universal.

\section{Acknowledgments}

I thank the support of CONACYT M\'exico, SNI M\'exico, EDI IPN, COFAA IPN, and Research Projects IPN SIP-20140832 and IPN SIP-20144150.

\end{document}